\documentclass[conference]{IEEEtran}
\IEEEoverridecommandlockouts
\usepackage{cite}
\usepackage{mathtools}
\usepackage{url}
\usepackage{blindtext}
\usepackage{wrapfig}
\usepackage{enumerate}
\usepackage{footnote}
\usepackage{amsmath,amssymb,amsfonts}
\usepackage{graphicx}
\usepackage{textcomp}
\usepackage{placeins}
\usepackage{xcolor}
\usepackage{verbatim}
\usepackage{algpseudocode,algorithm,algorithmicx}
\usepackage{footnote}
\usepackage{mathtools}
\usepackage{booktabs}
\usepackage{multirow}
\usepackage{siunitx}
\usepackage{float}
\usepackage{placeins}

\DeclarePairedDelimiter\floor{\lfloor}{\rfloor}

\algrenewcommand\algorithmicrequire{\textbf{Precondition:}}
\algrenewcommand\algorithmicensure{\textbf{Postcondition:}}

\makeatletter
\newenvironment{PurelyF}[1][htb]{%
    \renewcommand{\ALG@name}{Naive purely forward algorithm}
    
   \begin{algorithm}[#1]%
  }{\end{algorithm}}
\makeatother



\makeatletter
\newenvironment{NaiveForceReach}[1][htb]{%
    \renewcommand{\ALG@name}{NAIVE\_FORCE\_REACH}
    
   \begin{algorithm}[#1]%
  }{\end{algorithm}}
\makeatother

\makeatletter
\newenvironment{Auxiliary_funs}[1][htb]{%
    \renewcommand{\ALG@name}{Multiple perspective algorithm: auxiliary functions}
    
   \begin{algorithm}[#1]%
  }{\end{algorithm}}
\makeatother

\makeatletter
\newenvironment{OptAlgo}[1][htb]{%
    \renewcommand{\ALG@name}{Multiple perspective algorithm}
    
   \begin{algorithm}[#1]%
  }{\end{algorithm}}
\makeatother

\makeatletter
\newenvironment{ReachGame}[1][htb]{%
    \renewcommand{\ALG@name}{Improved purely backward algorithm}
    
   \begin{algorithm}[#1]%
  }{\end{algorithm}}
\makeatother

\makeatletter
\newenvironment{Improved_algo_auxiliary_funs}[1][htb]{%
    \renewcommand{\ALG@name}{Improved purely backward algorithm: auxiliary functions}
    
   \begin{algorithm}[#1]%
  }{\end{algorithm}}
\makeatother

\def\BibTeX{{\rm B\kern-.05em{\sc i\kern-.025em b}\kern-.08em
    T\kern-.1667em\lower.7ex\hbox{E}\kern-.125emX}}
\begin{document}

\title{Games on Graphs: A Time-Efficient Algorithm for Solving Finite Reachability and Safety Games}
\author{
\IEEEauthorblockN{Christian Giannetti}
\IEEEauthorblockA{Sapienza University of Rome}
\IEEEauthorblockA{giannetti.1904342@studenti.uniroma1.it}
}

\maketitle

\begin{abstract}
In recent years, there has been a growing interest in games on graphs within the research community, fueled by their relevance in applications such as economics, politics, and epidemiology. This paper aims to comprehensively detail the design decisions involved in developing a time-efficient algorithm for solving finite reachability and safety games on graphs. The primary contribution of this work is the introduction of a novel algorithm that effectively addresses both reachability and safety games by exploiting their inherent duality. The performance of the proposed algorithm is rigorously evaluated against traditional methods using a randomized testing framework. The paper is organized as follows: first, we provide the reader with a theoretical overview of reachability and safety games, followed by an in-depth discussion on the construction of the playing arena. A formal definition of reachability and safety games and a review of traditional algorithms for their resolution are then presented. Subsequently, the multiple-perspective algorithm is introduced along with its optimizations. The paper concludes with an extensive set of experiments and a comprehensive discussion of their results.
\end{abstract}
\section{Introduction}
The present section includes a theoretical compendium that introduces and rigorously formalizes several notions, from the definition of deterministic turn-based games to the theorem of determinacy of reachability games with its corollaries and the theorem of duality between reachability and safety games. The information contained in this section has been extracted from multiple sources \cite{gradel2002automata,thomas2002infinite,mavronicolas2008algorithmic,zimmermann2014infinite}, aggregated, and reviewed to ensure a self-contained theoretical summary that supports the reader's understanding without compromising the rigor required by the discussion.
\subsection{Deterministic turn-based games}
\textbf{Fundamental definitions} \indent
In deterministic games, a game graph is represented by the structure $G = (V,E,V_0,V_1)$, where $G$ consists of a directed graph $(V, E)$ and a partition $(V_0, V_1)$ of its set of vertices $V = V_0 \cup V_1$. The vertices and edges of a game graph are referred to as \textit{positions} and \textit{moves}, respectively. Moreover, $E^+(v)$ represents the set of successors of a position $v \in V$ (i.e., the set of nodes $E^+(v) = \{s : (v, s) \in E\}$), and $E^-(v)$ denotes the set of predecessors of a position $v \in V$ (i.e., the set of nodes $E^-(v) = \{p : (p, v) \in E\}$). A typical play on a game graph involves two players, Player 0 and Player 1, who establish a path by moving a token along the graph's edges. At the start of the play, the token is placed on a specific position.
\newline \newline \newline
\indent \textbf{Notion of turn-based game} \indent
In the context of our discussion, we will refer to the partition $(V_0, V_1)$ as the \textit{turn partition}. The concept of \textit{turn-based} play is determined by the player who has control over the node where the token is located. Specifically, if the current position $v$ is in the set $V_0$, then it is Player 0's turn to move the token to a successor $s \in E^+(v)$; conversely, if $v \in V_1$, then it is Player 1's turn to move. Therefore, a play that commences from a specified position $v_0 \in V$ in graph G can be represented as an infinite sequence of positions $\pi = v_0, v_1,\; \dotsc\;, v_n$, outlining a path on the graph $G$. Furthermore, we define an \textit{initial play} as a prefix of a play.\newline \newline \indent
\textbf{Definition of strategy} \indent 
A \textit{strategy} for Player 0 is defined as a function $\sigma : V ^{\ast} V_0 \mapsto V$ that maps every initial play $v_0, v_1,\; \dotsc\;, v_n$ terminating in a position $v_n \in V_0$ to a successor position $v_{n+1} \in E^+(v)$. A play $\pi = v_0 , v_1 ,\; \dotsc\;$ follows the strategy $\sigma$ if $v_{n+1} = \sigma(v_0, \;\dotsc\;, v_n)$ for all $n$ with $v_n \in V_0$. Similarly, a strategy is defined for Player 1. Once a starting position is determined, any pair of strategies $(\sigma, \tau)$ for Player 0 and Player 1, respectively, determines a play. The unique play that follows both $\sigma$ and $\tau$ is called the \textit{outcome} of the two strategies, denoted by $\sigma ^\wedge \tau$. \newline \newline \indent
\textbf{Definition of winning condition} \indent 
A winning condition on the set $G$ consists of a set $W \subseteq V^\omega$ of plays. If a play $\pi$ is in $W$, then Player 0 wins; otherwise, Player 1 is the winner. A strategy is considered winning for a player if all plays that result from the strategy are winning. In conclusion, a game $G = (G, W)$ is represented by defining a game graph and a winning condition.
\subsection{Reachability and safety games}\indent 
We define a reachability condition using a set $F \subseteq V$ of \textit{target} positions, which describes the winning condition as follows:
\begin{align}
\{v_0, v_1, \;\dotsc, \;: v_n \in\;F\;for\;some\;n\;\geq 0\} \nonumber
\end{align} \indent 
Similarly, we denote a safety condition using a set $F \subseteq V$ of \textit{safe} positions, which describes the winning condition as:
\begin{align}
\{v_0, v_1, \;\dotsc, \;: v_n \in\;F\;for\;all\;n\;\geq 0\} \nonumber
\end{align}\indent 
For the sake of completeness, we define the concept of infinitary conditions by considering the set of elements that occur an infinite number of times in a sequence:
\begin{align}
Inf(v_0, v_1,v_2,\;\dotsc\;) \;: \{ v = v_n\;\text{ \textit{for infinitely many} } n\geq 0\} \nonumber
\end{align}  \indent 
It is worth noting, however, that only a finite number of moves is necessary to determine whether a reachability condition is met or a safety condition is violated.
\newline \newline \indent 
\textbf{Determinacy of reachability games} \indent 
When concluding if a game is determined or not, we are essentially trying to establish whether, given a class $\Gamma$ of games, for any $G \in \Gamma$, either Player 0 or Player 1 possesses a winning strategy.
To provide a comprehensive view, it is worth noting that an alternative approach involves investigating whether, in the game $G = (V, W)$, there exists a partition $V = W_0 \cup W_1$ such that Player 0 has a winning strategy from any position $v \in W_0 $, while Player 1 has a winning strategy from any position $v \in W_1$. In such a scenario, the sets $W_0$ and $W_1$ are referred to as the \textit{winning regions} of Player 0 and Player 1, respectively.
\newline \indent In this context, reachability games (i.e., games with reachability winning conditions) are determined, as stated by the following theorem:\newline \newline \indent \textit{\textbf{Theorem of determinacy of reachability games}} \indent
\textit{For every reachability game $G = (V, V_0 ,E ,F)$, there exists a partition $W_0 \cup W_1 = V$ such that Player 0 has a winning strategy from any starting position $v \in W_0$ and Player 1 has a winning strategy from any starting position $v \in W_1$}. \newline \newline \indent
\textit{\textbf{Theorem of determinacy of reachability games: proof}} This paragraph proves the determinacy theorem for reachability games on finite graphs. In this proof, we inductively construct a sequence of sets $((Attr^0_i(F))_{i\geq0}$ with the property that from any position within $Attr^0_i(F)$, Player 0 can guarantee reaching $F$ in a maximum of $i$ steps.
\begin{equation}    
\begin{split}
  & Attr^0_0(F) \coloneqq F; \\
  & Attr^0_{i+1}(F) \coloneqq Attr^0_i(F) \\ 
  & \cup \{v \in V_0: E^{+}(v) \cap Attr^0_i(F) \neq \emptyset \} \\
  & \cup \{v \in V_1: E^{+}(v) \subseteq Attr^0_i(F)\} \\ \nonumber
\end{split} 
\end{equation}
The sequence is increasing until it reaches a fixed point $Attr^0_{i}(F) = Attr^0_{i+1}(F)$ after at most $|V|$ many stages. We denote this fixed point by $Attr^0(F)$ and call it the $attractor$ of $F$ for Player 0. 
\newline We claim that:
\begin{equation}    
\begin{split}
  &W_0 \coloneqq Attr^0(F) \text{ is the winning region of Player 0 on $G$}  \\ 
  &W_1 \coloneqq V \backslash Attr^0(F) \text{ is the winning region of Player 1 on $G$.}\\ \nonumber
\end{split} 
\end{equation}
To demonstrate this, we define a function $rank : V \xrightarrow{} \omega \cup \{\infty\}$ that associates to every position $v \in V$ the stage at which it was included into the attractor, 
\begin{equation*}
rank(v) \triangleq 
\left\{
\begin{alignedat}{2}
 min\{i : v \in Attr^0_i(F)\}\;\;\;\; \text{if v} \in Attr^0(F)\\
\infty \hspace{95pt}\text{if v} \notin Attr^0(F)\\
\end{alignedat}
\right.
\end{equation*}
The function mentioned above has the following property: \newline \newline 
\textit{(i)} for every $v \in Attr^0(F)$, either
\begin{equation}    
\begin{split}
  & \bullet v \in F, \text{or} \\ 
  & \bullet v \in V_0 \backslash F \text{ and for some successor w} \in E^+(v),\\
  & rank(w) < rank(v), \text{or} \\
  & \bullet v \in V_1 \backslash F \text{ and for all successors w} \in E^+(v),\\
  & rank(w) < rank(v);\\ \nonumber
\end{split}
\end{equation} 
\textit{(ii)} $\text{for every v} \in V \backslash Attr^0(F)$, either
\begin{equation}    
\begin{split}
  & \bullet v \in V_0, \text{ and, for all successors } w \in E^+(v), \\ 
  & rank(w) = \infty, \text{ or} \\
  & \bullet v \in V_1, \text{ and, for some successor } w \in E^+(v), \\ 
  & rank(w) = \infty. \\ \nonumber
\end{split} 
\end{equation}
Accordingly, we can define a function $f : V_0 \xrightarrow[]{} V$ that selects for every position $v \in V_0$ a successor $f(v) \in E^{+}(v)$ such that $rank(f(v))<rank(v)$ whenever $v \in Attr^0(F) \backslash F$. \newline Then, we consider the reachability game $G$ with an arbitrary starting position $v_0 \in Attr^0(F)$, and let $\sigma: V ^{\ast} V_0 \xrightarrow[]{} V$ be the strategy for Player 0 that chooses for every initial play $\pi = v_0, v_1, \dotsc , v_l$ with $v_l \in V_0$ the successor $f(v_l)$. Then, any play $v_0,  v_1, v_2 , \dotsc$ that follows $\sigma$ must reach a position of $F$. Otherwise, we had a strictly decreasing sequence $$ rank(v_0) > rank(v_1) > rank(v_2) > \dotsc $$ of infinite length, which cannot happen. Thus, $\sigma$ is a winning strategy for Player 0 in $G$, $v_0$. \newline
Conversely, for Player 1, let $g : V_1 \xrightarrow[]{} V$ be a function that selects for every $v \in V_1$ a successor $g(v)$ such that $rank(g(v)) = \infty$ whenever $v \in V\backslash Attr^0(F)$. For the game $G$ starting at an arbitrary position $v_0 \in F \backslash Attr^0(F)$, define a strategy $\tau : V ^{\ast} V_1 \xrightarrow[]{} V$ for Player 1 by associating to every initial play $v_0, v_1, \dotsc v_{l}$ with $v_{l} \in V_1$ the successor $g(v_{l})$. Then, in any play $v_0, v_1, v_2, \dotsc$ following $\tau$, we have $rank(v_i) = \infty$ at all indices $i \geq 0$, which means that $F$ is never reached. Hence, $\tau$ is a winning strategy for Player 1 in the reachability game $G$, $v_0$. \newline \newline \indent
\textbf{Memoryless strategy} \indent 
A strategy $\sigma: V ^{\ast} \rightarrow V_0$ for a game $G$ with initial vertex $v_0$ is defined \textit{positional} or \textit{memoryless} if there exists a function $f : V_0 \rightarrow V$ such that $\sigma(v_0, v_1 , v_2 , \dotsc , v_n) = f(v_n)$ for all initial plays $v_0, v_1 , \dotsc , v_n$ with $v_n \in V_0$. The existence of such function $f$ induces a memoryless strategy.
\newline \indent
For a fixed game $G$, a memoryless strategy $f : V_0 \mapsto V$ induces a (proper) strategy $\sigma: V ^{\ast} V_0 \mapsto V$ for every game $G$ with initial position $v_0$ where $v_0 \in V$. Such a memoryless strategy is defined as \textit{uniformly winning} over a set of positions $U \subseteq V$ if the induced proper strategy is winning in every game $G$ beginning at a position in $U$.
\newline \newline \indent
\textbf{First corollary of reachability games' theorem of determinacy} \newline \indent \textit{For every reachability game $G$, the set $V$ of positions can be partitioned into $W_0 \cup W_1 = V$ such that Player 0 has a uniform memoryless winning strategy over $W_0$ and Player 1 has a uniform memoryless winning strategy over $W_1$. The winning regions, jointly with the memoryless strategies, can be computed in time} $\mathcal{O}(|V|+|E|)$.
\newline \newline \indent 
\textbf{Duality of reachability and safety games} \indent 
Every result drawn for reachability games applies readily to safety games since reachability games and safety games are dual. Specifically, given a fixed game $G$, it is possible to solve the game from both the players' perspectives. \newline Formally, we first define the following propositions: \noindent
\begin{equation}    
\begin{split}
  & P_1 = \text{Player 0 has a winning strategy in the safety game } 
  \\&(V, V_0, V_1, E, F) \nonumber
  \\& P_2 = \text{Player 1 has a winning strategy in the reachability game} \nonumber
  \\&(V, V_1, V_0, F_0 = V \backslash F), \text{where the players' roles are switched.} \nonumber
\end{split} 
\end{equation}
Then, we observe that the following condition applies:
\begin{equation}
P_1 \iff P_2 \nonumber
\end{equation} \indent
\textbf{Second corollary of reachability games' theorem of determinacy} \newline \indent \textit{For every safety game $G$, the set $V$ of positions can be partitioned into $W_0 \cup W_1 = V$ such that Player 0 has a uniform memoryless winning strategy over $W_0$ and Player 1 has a uniform memoryless winning strategy over $W_1$. The winning regions, jointly with the memoryless strategies, can be computed in time} $\mathcal{O}(|V|+|E|)$.

\subsection{Representing the graph: data structures employed}
This paragraph describes the data structure employed to represent the graph. The function that builds the graph takes as input:
\begin{itemize}
    \item The nodes controlled by the reachability player, labeled with an integer. 
    \item The nodes controlled by the safety player, labeled with an integer. 
    \item The edges between the nodes present in the graph. 
\end{itemize}
Given these inputs, the function generates two graphs: the straight and transpose graphs. The latter is employed in the optimized versions of the presented algorithms to limit the computational resources in specific scenarios and decrease the time required to solve the game. At the implementation level, we model each graph through a dictionary that associates an integer representing the node's label with a set of integers that include all the nodes' neighbors. In other terms, each integer in the set represents a graph node. We detail the reasons why exploiting the transpose graph in section \ref{Optimizations} is convenient.
\subsection{Random graph generation: automatic experiments}
In order to facilitate and speed up the testing procedure, we provide a function that automatically generates a random graph. We have integrated this function into the code, allowing the user to generate large graphs quickly. Precisely, the function takes as input the following parameters:
\begin{itemize}
    \item \textit{Number of nodes}: an integer that specifies the number of nodes.
    \item \textit{Number of edges}: an integer that specifies the number of edges. 
    \item \textit{Self-loops}: a Boolean that regulates the presence of self-loops in the graph.
    \item \textit{Isolated nodes}: a Boolean that regulates the presence of isolated nodes in the graph.
\end{itemize}
A crucial matter during the testing phase is guaranteeing repeatable and unbiased tests. For this reason, we also provide a function responsible for executing a repeatable, automatic, and unbiased battery of tests. Specifically, the function takes the following inputs:
\begin{enumerate}[a]
    \item $:=$ \textit{num\_nodes\_min}
    \item $:=$ \textit{num\_nodes\_max}
    \item $:=$ \textit{avg\_edges\_per\_node\_min}
    \item $:=$ \textit{avg\_edges\_per\_node\_max}
    \item $:=$ \textit{target\_safe\_ratio\_min}
    \item $:=$ \textit{target\_safe\_ratio\_max}
\end{enumerate}\noindent
Given the above inputs, the function generates multiple graphs and uses them as arenas to benchmark multiple algorithms. Specifically, for each experiment requested by the user, it computes the following parameters:
\begin{itemize}
    \item \textit{num$\_$nodes:= }$randInt(a, b)$. This parameter specifies the number of nodes in the graph.
    \item \textit{num$\_$edges:= }$\floor{num$\_$nodes * randFloat(c, d)}$. This parameter specifies the number of edges of the graph.
    \item \textit{target$\_$safe$\_$size:= }$\floor{num$\_$nodes * randFloat(e, f)}$ \newline This parameter defines the ratio that regulates the dimension of the target set for the safety player with respect to the total number of nodes in the graph (e.g., a ratio of 0.7 indicates that the target set has a dimension equal to the 70\% of the total number of nodes in the graph).
\end{itemize}
\subsection{Random graph generation: avoiding isolated nodes}
The randomly generated graph employed in the experimental phase may contain isolated nodes. Their presence increases the computational cost of the algorithm's execution but does not add any informative content to the experiments. Since this phenomenon becomes significant in large graphs, we propose a method that, given a randomly generated disconnected graph, returns a connected graph that is finally used to perform the desired experiment. \newline We explain in the following lines the simple logic underpinning the method. We first generate an $N\text{x}N$ adjacency matrix whose element $(i,j) = 1$ if the edge $(i,j)$ is present in the graph; otherwise, $(i,j) = 0$. Then, for each $(i,j)$, we add 1 with a probability such that the expected value of the edges' number equals an integer parameter taken as input, denoted as $E$. If the option $no\_isolated$ is set to $true$, we iterate through the matrix to verify if it contains isolated nodes. Specifically, for each row of the matrix, we check that all the row's elements equal $0$. If this is the case, we randomly select an integer $j$ to determine which column will be modified. Then, we flip a fair coin to decide whether to add the edge $(i,j)$ or $(j, i)$. This process is repeated until the graph has no longer isolated nodes.
\section{From theory to practice: problem formulation from different perspectives}
\subsection{Safety-reachability duality: \newline Reachability problem formulation}
Given a game, seen from the reachability player point of view, it is possible to decompose the set $Force_R(X)$ as follows:
\begin{align}
& Reach\_comp(X) =  \{v \in V_R: E(v) \cap X \neq \emptyset \} \nonumber \\
& Safety\_comp(X) = \{v \in V_S: E(v) \subseteq X \} \label{Form1} \\ 
& Force_R(X) = Reach\_comp(X) \cup Safety\_comp(X) \nonumber
\end{align} 
We remark that in a safety game, player's roles are inverted. In this scenario, the reachability component $Reach\_comp(X)$ includes all the nodes whose successors are in the safety player's winning set, i.e., the nodes the reachability player cannot avoid entering the region $X$. On the other hand, the safety component $Safety\_comp(X)$ includes all the nodes with at least a successor in the safety player's winning set, i.e., the nodes for which the safety player has a move to enter in the region $X$. It follows that the set $Force_S(X)$ contains the nodes for which the safety player can enforce the token to remain in his winning set forever.
\subsection{Safety-reachability duality: naive purely forward algorithm}
As previously pointed out, the algorithm below computes the set of nodes from which the safety player can remain in the target set forever. Given as input the target set, a set of candidate safe nodes, the algorithm removes at each iteration those nodes from which the reachability player has a move to enter his target set (i.e., the unsafe nodes for the safety player). In doing so, as the game progresses, each unsafe node is removed from the target set, and the algorithm finally determines the winning set for the safety player. In this respect, we denote this paradigm to solve the game as \textit{proceeding forward}.
\begin{PurelyF}
\caption{}
\begin{algorithmic}[1]
\Statex{\textbf{Input:}}
\Statex{Graph $G(V, E)$: The graph representing the arena.}
\Statex{Target\_safe: The target set for the safety player.}
\Statex{\textbf{Begin:}}
\State{Win = Target\_safe}
\While{(Win $\neq$ (Win $\cap$ force(Win)):}
    \State{Win = Win $\cap$ force(Win)}
\EndWhile
\State{\textbf{return} Win}
\end{algorithmic}
\end{PurelyF}
\noindent
\section{Optimizations}
\label{Optimizations}
The following sections discuss in detail the algorithmic optimizations proposed and provide their theoretical justifications. We remark that all the proposed optimizations are not mutually exclusive; thus, it is possible to use them simultaneously.

\subsection{Transpose graph optimization}
\label{TGO}
The following analysis aims to provide the reader with the intuition that it is convenient to employ the transpose graph in solving reachability or safety games. We provide an example using formulation $(1)$ for clarity. To compute $Reach\_comp(X)$, that is the component of $Force_R(X)$ related to the reachability player; it is necessary to iterate through the list of all the nodes in the straight graph and verify that from the node $v$, it is possible to reach the region $X$. We observe that, by construction, the target set is always a subset of the complete set of the graph nodes (i.e., the total number of nodes in the graphs is always greater or equal than the number of nodes belonging to the target set). Note that their difference, in terms of nodes' number, can also span several orders of magnitude. \hfill \break \indent
We provide an alternative method to reduce the computational cost of computing $Reach\_comp(X)$. Specifically, when computing nodes to add in $Reach\_comp(X)$, we would like to consider only those nodes in $V$ with an edge to a node that belongs to the target set $X$. This observation indicates that it is beneficial to employ a transpose graph to compute these nodes directly. More precisely, instead of considering each node $v \subseteq V$ from which it is possible to reach the target set $X$ in the straight graph, it is much more convenient to examine each node $v \subseteq V$ reached from a node $x \subseteq X$ in the transpose graph. This optimization exploits the fact that the target set's size, on average, is smaller than the size of $V$. It follows that the optimization is particularly effective in the algorithm's first iterations because the target set's size increases at each consecutive iteration.\footnote{This observation holds if the game is solved from the reachability player's point of view. Conversely, if the game is solved from the safety player's point of view, the winning set's size decreases at each iteration.} Specifically, the only scenario in which the two approaches have exactly the same computational complexity is the one in which the graph $G(V, E)$ is complete (i.e., every node $v \subseteq V$ is connected to all the remaining graph nodes). In this latter case, the computational complexity of both approaches becomes quadratic in the number of nodes. However, such a scenario is improbable in the case of randomly generated graphs, which are generally sparse. \hfill \break \indent
We note that this optimization becomes less and less effective the denser the graph, but it does increase the algorithm's scalability on average.

\subsection{Transpose graph optimization: \newline simultaneous usage of multiple graphs}
Given a graph $G$ and an input target set $X$, we define the neighborhood of the target set in the transpose graph $G^T$ as follows: 
\begin{align}
& N_{G^T} := \{ x \in X\;\vert \;(x \subseteq Neighbors(G^T,x))\} \nonumber
\end{align}
Specifically, $N_{G^T}$ indicates the set of nodes reachable from a node belonging to the target set through a single edge in the transpose graph.
\hfill \break \indent
\hfill \break \indent
To provide the reader with an intuition of this hybrid approach, we take a reachability game as an example. Regarding formulation \ref{Form1}, the following lines investigate the conditions the nodes have to satisfy to be included in a particular component of $Force_R(X)$, either $Reach\_comp(X)$ or $Safety\_comp(X)$.
\hfill \break \indent
\hfill \break \indent
Concerning $Reach\_comp(X)$, the set will include all the nodes $v \subseteq V$ that satisfy the following conditions:
\begin{enumerate}[a)]
	\item The node v is controlled by the reachability player \newline (i.e., $v \cap R$).
	\item The node v belongs to the set $N_{G^T}$ (i.e., $v \in N_{G^T}$).
\end{enumerate}
As pointed out in section \ref{TGO}, it is convenient to employ the transpose graph instead of the straight graph to verify that the condition \textit{b)} is fulfilled. This design choice has the advantage of reducing the computations required for this last check. \newline
\hfill \break \indent
On the other hand, $Safety\_comp(X)$ will include all the nodes $v \subseteq V$ that satisfy the following conditions:
\begin{enumerate}[a)]
	\item The node v is controlled by the safety player \newline (i.e., $v \cap S$).
	\item The node v belongs to the set $N_{G^T}$ (i.e., $v \in N_{G^T}$).
	\item All the outgoing edges of the node $v \subseteq V$ lead to nodes contained in the target set $X$ 
	(i.e., $v \in V \; \vert \; E(v) \subseteq X$). 
\end{enumerate}
As seen above, the computation of $Safety\_comp(X)$ imposes to verify an additional condition, for which it is strictly necessary to use the straight graph. Hence, we propose a hybrid approach. We employ the transpose graph to check the fulfillment of condition \textit{b)}, while we use the straight graph to check if condition \textit{c)} is verified. We act this way to optimize the process to the maximum extent possible.
\subsection{Transpose graph optimization: applicative example}
\begin{figure}[htbp]
    \centering
    {\includegraphics[width=1\linewidth]{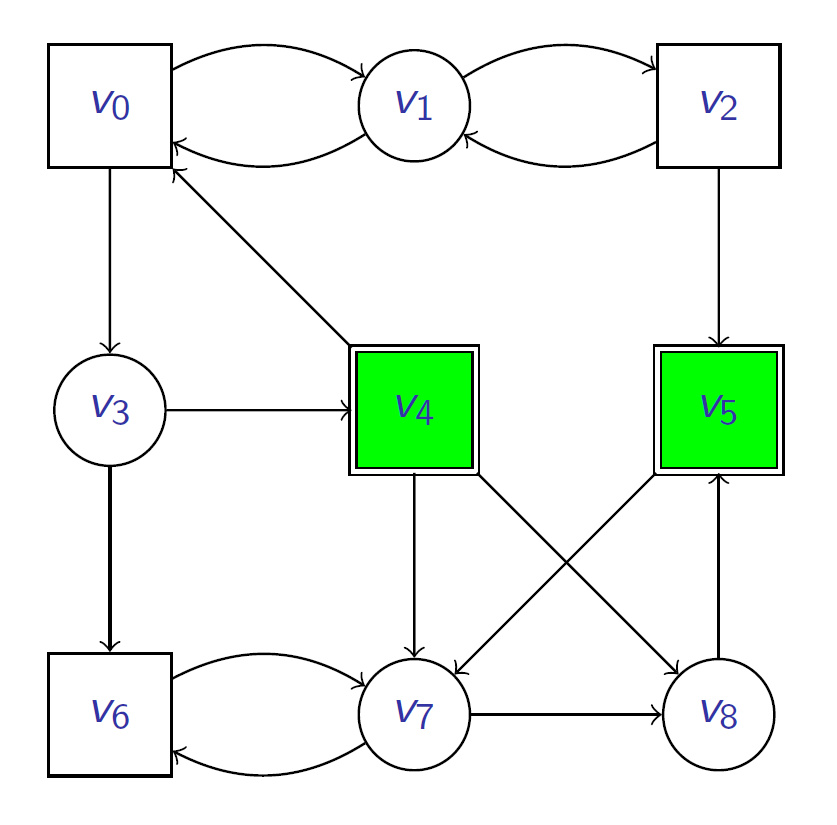}}
    \caption{Graphical representation of the given arena. The round-shaped nodes are the nodes controlled by the reachability player, while the square-shaped nodes indicate the nodes controlled by the safety player. Lastly, the region colored in green identifies the reachability of the player's target set.}
    \label{fig:im1}
\end{figure}

We show an applicative example to provide the reader with proof of the effectiveness of such optimization. Given the arena depicted in figure \ref{fig:im1}, it is needed to compute the nodes that satisfy the conditions \textit{a)} and \textit{b)} with the least possible computations. However, while employing the transpose graph generally decreases the computational cost of verifying whether condition \textit{b)} is fulfilled, the optimization does not affect the computational cost of checking the condition \textit{a)}. With reference to the figure \ref{fig:im1}, we distinguish the following scenarios depending on which data structure is used to execute the required computations:
\begin{enumerate}
    \item \textit{Straight graph usage: } When using the straight graph to compute which nodes satisfy the condition \textit{b)}, we will have the following scenario. The algorithm iterates on all the nodes in the graph controlled by the reachability player and, for each node, checks if the node has an edge to a node contained in the target set (i.e., the region colored in green). Thus, the set of processed nodes will be $V_p =  \{v_1, v_3, v_7, v_8\}$.
    \item \textit{Transpose graph usage: } On the contrary, when using the transpose graph, the scenario evolves as follows: the algorithm iterates only on the nodes reachable from the target set in the transpose graph. Hence, the set of nodes that must be processed will be restricted to $V_p' = \{v_3, v_8\}$.
\end{enumerate}
Regardless of the used data structure, $Reach\_comp$ will contain the same nodes, i.e., $Reach\_comp=\{v_3, v_8\}$.
\subsection{Optimizations for reachability games}
The following paragraphs discuss the algorithmic optimizations designed for solving reachability games more efficiently. Integrating these optimizations into the \textit{multiple-perspective} algorithm has improved the efficiency of the \textit{step backward}, which corresponds to solve the game from the reachability player's point of view.
\subsection{Optimizations for reachability games:\newline Current set optimization}
\begin{figure}[htbp]
    \centering
    {\includegraphics[width=1\linewidth]{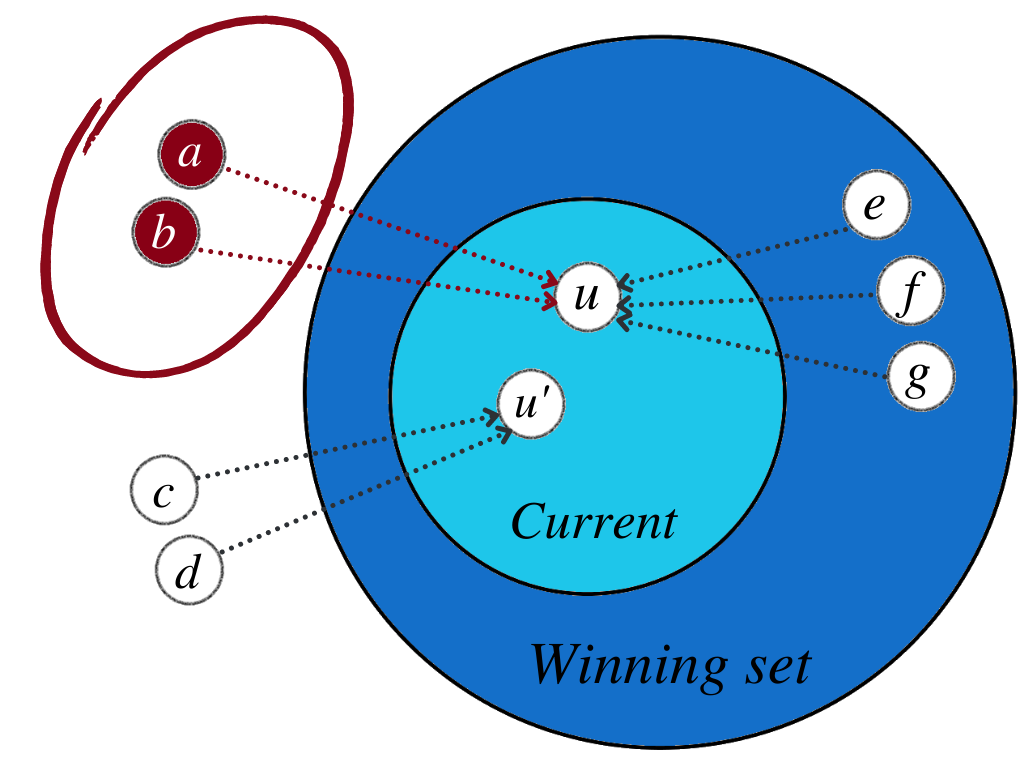}}
    \caption{Graphical representation of the functioning of the Current set optimization.}
    \label{fig:im2}
\end{figure}

Bearing in mind the formulation $(\ref{Form1})$, we define the \textit{Current set} as \textit{the $Force_{R}$ computed in the previous iteration of the algorithm}. Concisely, it denotes the set of nodes added in the last iteration to the reachability player's winning set. This observation demonstrates that the Current set is generally a subset of the target set.
This other optimization improves the computation efficiency of $Reach\_comp$ and can be combined efficaciously with the transpose graph optimization. Precisely, this optimization consists of verifying the fulfillment of the conditions \textit{a)} and \textit{b)} exclusively for the nodes reachable from the Current set instead of verifying whether these conditions hold for the set of nodes reachable from the entire target set.\footnote{Please note that we are reasoning by using the transpose graph. Hence, instead of stating that a node $v$ \textit{can reach} the target set $X$, we state that a node $v$ \textit{is reachable from} the target set $X$.} In fact, we observe that the nodes candidates to enter the target set are only those that have not been considered yet (i.e., the nodes reachable from the target set through a single edge). This occurs because the nodes added in the previous iteration (i.e., the nodes reachable from the target set through two edges) fall necessarily within the following cases:
\begin{enumerate}
    \item \textit{The node is not reachable from the target set with a single edge: } The node has not been added to the target set in the previous iteration.
    \item \textit{The node is not reachable from the target set through two edges: } The node must be necessarily added to the force set (and thus to the target set) in the previous iteration. Since the graph edges do not change between the iterations, it is not possible that a node was not reachable from the target set through a single edge in the previous iteration, but that becomes reachable in the current iteration.
\end{enumerate}
For the reasons presented in this paragraph, it is convenient to verify whether the conditions \textit{a)} and \textit{b)} hold \textit{only for the nodes contained in the Current set}.\newline
Figure \ref{fig:im2} provides a graphical explanation of the optimization presented above. With reference to the figure, suppose the algorithm is processing the node \textit{u}. The only candidate to the entrance in the force set we would like to consider are the nodes \textit{a} and \textit{b}, circled in red. 
The reasons are the following. The nodes \textit{c} and \textit{d} cannot reach the node \textit{u}, so they will be taken into account when processing the node \textit{u'}. On the other hand, the nodes \textit{e, f, and g} could be considered in the current passage, but they are already in the winning set; hence, adding them to the force set would not be beneficial. More precisely, it would be detrimental because the computational cost of performing the union would increase when considering some nodes already in the winning set.
\subsection{Optimizations for reachability games:\newline Processed list optimization}
Keeping in mind the formulation $(\ref{Form1})$,  we first introduce the following points:
\begin{itemize}
    \item We suppose to solve the game as a safety game.\footnote{We are considering this perspective to align with the provided code. The optimization in question is incorporated into the "multiple-perspective algorithm," which initiates the game-solving process from the safety player's standpoint.} Hence, we denote the winning set for the safety player as $W$, and its complement, the losing set, as $L$.
    \item In the present paragraph, we will indicate with $V_S$ the \textit{set nodes controlled by the safety player} and with \textit{SCL} the \textit{set of nodes controlled by the safety player that are solely connected with nodes belonging to} $L$.
\end{itemize}
The present optimization is related to the computation of $Safety\_comp$ in the function \textit{FORCE$\_$SAFE$\_$BACKWARD} of the \textit{multiple-perspective} algorithm. The function aims to compute the set \textit{SCL}. \newline The first step is to verify whether the conditions \textit{a)} and \textit{b)} hold. Instead of examining the whole set $V_S$, we examine only the nodes that can reach a losing node through a single edge by exploiting the transpose graph, as explained in the previous paragraphs.
At this point, it is necessary to verify whether the nodes considered satisfy the condition \textit{c)}, i.e., to identify all the nodes whose outgoing edges lead to nodes belonging to $L$. Specifically, we apply the following procedure. \newline For each node $u \in L$, we examine all the nodes controlled by the safety player solely connected with $u \in L$ or with another node $u' \in L$. We conduct this check by iterating through the set $V_S$. However, a node $v$ may be encountered, firstly, as a neighbor of the node $u$ and, secondly, of the node $u'$. Since the result of the operations carried out will be independent of the fact that the node $v$ is a neighbor either of the node $u$ or the node $u'$, we insert each considered node into \textit{Processed}, a list that contains the processed nodes. In doing so, once a node has already been examined as the neighbor of a node $u$, it will not be considered again if it is encountered as the neighbor of another node $u'$.\newline
\subsection{Safety-reachability duality: \newline improved \textit{purely backward} algorithm}
The algorithm presented in this paragraph follows strictly the logic of the \textit{purely backward} algorithm, but it integrates the optimizations discussed above to improve its performance. This optimized algorithm version is implemented for reachability games in the code provided. However, the algorithm can also solve safety games by simply returning the complement of the computed winning set. We also provide below the code of the function \textit{NAIVE\_FORCE\_REACH}, which is the main target of most of the optimizations. We encourage the reader to compare this function with the optimized \textit{FORCE\_REACH} integrated into the \textit{improved backward algorithm} to understand why the presented optimizations are particularly effective fully.
\subsection{Optimizations for reachability games:\newline \textit{multiple-perspective} algorithm}
The present algorithm combines the two logics underpinning the algorithms previously presented: \textit{pure forward} and \textit{improved pure backward}. The fundamental idea underlying the design of the present algorithm is to take the best from both approaches. Precisely, at each iteration, the algorithm determines whether it is convenient to tackle the problem from either the reachability or safety player's point of view, depending on a heuristic based on the winning set's size.
As a design choice, the algorithm starts to solve the game as a safety game, thus taking the target set for the safety player as input. At each iteration, it determines if it is convenient to proceed \textit{backward} (i.e., solving the problem from the reachability player's point of view) or \textit{forward} (i.e., solving the problem from the safety player's point of view). Once the safety player's winning set for the current iteration has been computed, the algorithm automatically computes the other player's target set, which is symmetrically named as the \textit{losing set}, finally allowing to switch the point of view, if needed. 
\subsection{Multiple-perspective algorithm: the employed heuristic}
\label{heuristic}
Since the complexity of computing either the sets of both players is dominated by the number of nodes in the winning set, the algorithm determines whether it is convenient to perform the \textit{forward} or \textit{backward} steps depending on the winning set's size. More precisely, whenever the winning set's size is less than or equal to half the number of the nodes in $V$, performing the forward step is convenient; otherwise, executing the backward step is convenient. Note that the actual complexity of each step depends on the degree of the nodes involved in the computation. Hence, this criterion is not an accurate indicator, but it is just used as a heuristic.
Lastly, emphasis must be placed on the following point: \textit{taking a step instead of another does not undermine the correctness of the algorithm}. To provide the reader with a less formal but immediate intuition, we point out that changing the point of view at each iteration can be viewed as generating at each iteration a new game, in which the target set given in input is the winning set returned by the step executed at the previous iteration. In this respect, each iteration is independent of the previous one; hence, switching between the two points of view as the game progresses does not affect the algorithm's correctness.
\section{Experimental phase}
Concisely, the experimental phase is structured as follows. In the first place, the script automatically generates an arena (i.e., a graph), where it launches multiple algorithms that solve the same game. At the end of each execution, the results are collected to provide a set of statistics to the user. Precisely, we benchmark the proposed algorithms by measuring the time required to solve the same game. Concerning the experiments' structure, we have opted not to include the improved version of the backward algorithm in the proposed battery of tests. The reason is the following. The design of the improved algorithm to solve the reachability game has been undoubtedly propaedeutic to the design of the \textit{multiple-perspective} algorithm. However, we have preferred to evaluate the combined action of the multiple optimization techniques employed rather than testing the effects of every single optimization in a wide range of scenarios. This design choice lies in the fact that all the optimizations proposed are not mutually exclusive (i.e., using a technique does not preclude applying another technique). Specifically, each experiment envisages a run of the following algorithms:
\begin{enumerate}[a)]
    \item \textit{Naive purely forward algorithm}
    \item \textit{Naive purely backward algorithm}
    \item \textit{Multiple-perspective algorithm}
\end{enumerate}
Furthermore, it is valuable to contrast the \textit{multiple-perspective} algorithm with the naive versions of algorithms designed for solving reachability and safety games. This comparison offers a succinct but pragmatic demonstration of the efficacy of the proposed approach. Given that all the proposed algorithms, including the naive versions, are assured to resolve reachability and safety games, we evaluate them based on the time needed to solve the same game.

\subsection{Presentation of the results}
The present subsection provides a concise and practical overview of the different algorithms' performance during the benchmark process. Before presenting the results, we point out a fundamental design choice regarding the evaluation method: \textit{all the algorithms have provided the same inputs regardless of their inner functioning}. Specifically, to evaluate the effectiveness of employing a more sophisticated approach, the transpose graph is not given as input to the multiple-perspective algorithm, which has to generate it independently. Please note that \textit{the time required by the algorithm to solve the game also includes the time required for generating the transpose graph}. This design choice concerning the evaluation method has been made for two reasons. Firstly, to obtain unbiased results and, secondly, to determine whether it is convenient to use the multiple perspective algorithm, although it has to autonomously generate at runtime the transpose graph, which is indeed a computationally expensive procedure when handling massive graphs. In other terms, we aim to verify whether the multiple-perspective algorithm performs well, at least as its naive counterparts, despite the additional burden of generating the transpose graph. \newline
Since the threshold employed to regulate the switch of point of view has been set to half the number of nodes, we aimed to investigate which cases the algorithm performs statistically better than its naive counterparts. In the items below, we briefly summarize the content of the tables that report the obtained results.
\begin{enumerate}
    \item Table \ref{tab:01_05} presents the results for games on graphs with a number of nodes ranging from 100 and 1000 and with a target\_safe\_ratio randomly ranging between 0.01 and 0.5.
    \item Table \ref{tab:05_10} shows the results for games on graphs with a number of nodes ranging from 100 to 1000 and with a target\_safe\_ratio was randomly ranging between 0.5 and 0.1.
    \item Table \ref{tab:100_1000_nodes_nb} illustrates the results for games on graphs with a number of nodes ranging from 100 to 1000 and with a target\_safe\_ratio randomly ranging between 0.01 and 1.0.
    \item Table \ref{tab:5000_6000_nodes} presents the results for games on graphs with a number of nodes ranging from 5000 and 6500 and with a target\_safe\_ratio randomly ranging between 0.01 and 1.0.
\end{enumerate}
The experiments performed under settings $(1)$ and $(2)$ aimed to verify the performances of the multiple perspective algorithm when the algorithm was forced to perform only forward steps (case 1) and when the algorithm was obliged to perform at least one backward step (case 2). On the other hand, the experiments conducted in scenarios $(3)$ and $(4)$ intended to examine the algorithm's performance under general settings to test whether applying the multiple perspective algorithm was statistically convenient in medium-size and considerably large graphs, respectively.
\subsection{Discussion of the results}
With reference to the presented results, it is possible to make the following observations:
\begin{itemize}
    \item Under the settings (1) [Table \ref{tab:01_05}], the multiple-perspective algorithm performs statistically better than both the forward and backward algorithms. This result is presumably related to the processed list optimization that enables the multiple-perspective algorithm to avoid performing the same operation several times. Please refer to section \ref{Optimizations} for further details.
    \item Under the settings (2) [Table \ref{tab:05_10}], the multiple-perspective algorithm performs statistically better than both the forward and backward algorithms eight times out of ten. On the other hand, in two cases, the forward algorithm performs better than the multiple-perspective algorithm. However, from the magnitude of the percentages, we can see that the difference in terms of performance in these last two cases is not particularly significant. We hypothesize that this last result is because the multiple-perspective algorithm has two additional costs when solving games. The first consists of generating the transpose graph, and the second consists of the cost of switching point of view (i.e., computing the complement of the winning set).\footnote{Please note that the algorithm can switch the point of view only once during a game, according to the chosen heuristic.} In other terms, the multiple-perspective algorithm is slower than its naive counterparts when the time required to compute the additional data structures exceeds the time saved through the carried-out optimizations.
    \item The experiments conducted under the settings (3) [Table \ref{tab:100_1000_nodes_nb}] and (4) [Table \ref{tab:5000_6000_nodes}] demonstrate that employing the multiple-perspective algorithm performs better than its naive counterparts on the average case. The cases in which the multiple-perspective algorithm is slower than the naive ones show that the difference in the required time is not particularly relevant. However, there are some cases in which the algorithm profoundly outperforms both the naive counterparts (e.g., Table \ref{tab:100_1000_nodes_nb} - experiment 1, Table \ref{tab:5000_6000_nodes} - experiment 10).
    \item From all the experiments conducted, it is possible to see that the multiple-perspective algorithm profoundly outperforms the backward algorithm in every conducted experiment. This result is plausibly related to the fact that most optimizations of the multiple-perspective algorithm have been designed to improve the algorithm's capabilities to solve reachability games; hence, the multiple-perspective algorithm is more efficient under these settings. \newline
    Moreover, we remark on another crucial factor. As a design choice, the game starts as a safety game; hence, it is given as input the target\_set for the safety player. In this respect, it is clear that the backward algorithm is disadvantaged because it has to compute the complement of the target\_set and then solve the game as a reachability game. Please note that the multiple perspective algorithm experiences the same disadvantage if, at the start, the size of the winning set is greater than half of the total number of nodes. However, it performs better than the backward algorithm thanks to the implemented optimizations.
\end{itemize}
\section{Conclusions}
This paper presented the design choices and implementation details of the multiple-perspective algorithm, a time-efficient approach for solving finite reachability and safety games that exploits their inherent duality. The algorithm's performance was rigorously evaluated using a randomized testing framework. The extensive experiments and detailed discussion of the results demonstrate the algorithm's superiority over traditional methods in terms of efficiency. In future work, we aim to extend the proposed algorithm to address more complex classes of games on graphs. This extension will involve adapting the algorithm to handle additional game dynamics and objectives, further broadening its applicability in more diverse and challenging scenarios.

\begin{samepage}
\begin{table*}[ht]
    \centering
    \begin{tabular}{SSSSSS} \toprule
    {$Experiment\_label$} & {$Total\_nodes$} & {$Total\_edges$} & {$Target\_nodes$} & {$Safety\_nodes$} & {$Reachability\_nodes$} \\ \midrule
    1  & 557 & 768 & 77 & 279 & 278\\
    2  & 148 & 413 & 72 & 74 & 74 \\
    3  & 656 & 1811 & 312 & 327 & 329\\
    4  & 910 & 4134 & 218 & 454 & 456\\
    5  & 800 & 2405 & 231 & 399 & 401\\
    6  & 260 & 345 & 86 & 130 & 130\\
    7  & 356 & 495 & 167 & 179 & 177 \\
    8  & 965 & 2656 & 198 & 482 & 483\\
    9  & 319 & 1422 & 139 & 159 & 160\\
    10 & 674 & 2250 & 291 & 337 & 337\\ \bottomrule
\end{tabular}
    \centering
    \begin{tabular}{SSSSSS} \\ \toprule
    {$Experiment\_label$} & {$FW\_time$} & {$BW\_time$} & {$MP\_time$} & {$Time\_saving\_wrt\_FW$} & {$Time\_saving\_wrt\_BW$}\\ \midrule
    1  & 0.0011 s & 0.0225 s & 0.0007 s & 31.68\% & 96.71\% \\
    2  & 0.0010s & 0.0018s & 0.0007s & 35.50\% & 62.76\% \\
    3  & 0.0026s & 0.0281s & 0.0025s & 2.87\% & 91.11\% \\
    4  & 0.0027s & 0.0557s & 0.0024s & 10.74\% & 95.66\% \\
    5  & 0.0027s & 0.0438s & 0.0023s & 13.72\% & 94.69\% \\
    6  & 0.0008 s & 0.0050s & 0.0004s & 48.05\% & 91.29\% \\
    7  & 0.0013s & 0.0068s & 0.0009s & 26.61\% & 86.32\% \\
    8  & 0.0032s & 0.0650s & 0.0024s & 26.21\% & 96.34\% \\
    9  & 0.0012s & 0.0059s & 0.0010s & 17.82\% & 83.77\% \\
    10 & 0.0031s & 0.0324s & 0.0024s & 21.25\% & 92.52\% \\ \bottomrule \\
\end{tabular}
\caption{First experiments battery. Number of nodes ranging between [100, 1000], target\_safe\_ratio ranging between [0.1-0.5].}
\label{tab:01_05}
\end{table*}

\begin{table*}[ht]
    \centering
    \begin{tabular}{SSSSSS} \toprule
    {$Experiment\_label$} & {$Total\_nodes$} & {$Total\_edges$} & {$Target\_nodes$} & {$Safety\_nodes$} & {$Reachability\_nodes$} \\ \midrule
    1  &220 &461 &219 &111 &109 \\
    2  &110 &512 &108 &55 &55 \\
    3  &911 &2140 &810 &455 &456 \\
    4  &130 &414 &127 &65 &65 \\
    5  &266 &598 &133 &134 &132 \\
    6  &116 &161 &96 &59 &57 \\
    7  &269 &629 &218 &135 &134 \\
    8  &191 &675 &173 &95 &96 \\
    9  &164 &493 &145 &82 &82 \\
    10 &148 &419 &145 &74 &74 \\ \bottomrule
\end{tabular}

    \centering
    \begin{tabular}{SSSSSS} \\ \toprule
    {$Experiment\_label$} & {$FW\_time$} & {$BW\_time$} & {$MP\_time$} & {$Time\_saving\_wrt\_FW$} & {$Time\_saving\_wrt\_BW$}\\ \midrule
    1  & 0.0013s & 0.0013s & 0.0010s & 18.79\% & 23.99\% \\
    2  & 0.0004s & 0.0004s & 0.0004s & 1.06\% & 12.34\% \\
    3  & 0.0130s & 0.0196s & 0.0143s & -10.33\% & 26.75\% \\
    4  & 0.0005s & 0.0007s & 0.0004s & 1.44\% & 32.25\% \\
    5  & 0.0007s & 0.0037s & 0.0006s & 3.96\% & 82.61\% \\
    6  & 0.0004s & 0.0007s & 0.0004s & 11.17\% & 46.40\% \\
    7  & 0.0012s & 0.0021s & 0.0015s & -17.69\% & 32.20\% \\
    8  & 0.0010s & 0.0014s & 0.0008s & 20.63\% & 43.66\% \\
    9  & 0.0009s & 0.0012s & 0.0008s & 10.43\% & 23.59\% \\
    10 & 0.0007s & 0.0005s & 0.0005s & 30.19\% & 4.15\% \\ \bottomrule \\
\end{tabular}
\caption{Second experiments battery. Number of nodes ranging between [100, 1000], target\_safe\_ratio ranging between [0.5,1.0].}
\label{tab:05_10}
\end{table*}
\end{samepage}

\begin{samepage}
\begin{table*}[ht]
    \centering
    \begin{tabular}{SSSSSS} \toprule
    {$Experiment\_label$} & {$Total\_nodes$} & {$Total\_edges$} & {$Target\_nodes$} & {$Safety\_nodes$} & {$Reachability\_nodes$} \\ \midrule
    1  &996 &2432 &23 &26 &970 \\
    2  &573 &1734 &119 &30 &543 \\
    3  &789 &2882 &218 &361 &428 \\
    4  &514 &914 &55 &112 &402 \\
    5  &529 &2517 &2 &196 &333 \\
    6  &174 &420 &49 &49 &125 \\
    7  &502 &2167 &77 &26 &476 \\
    8  &793 &2768 &357 &614 &179 \\
    9  &313 &489 &24 &9 &304 \\
    10 &203 &445 &196 &11 &192 \\ \bottomrule
\end{tabular}
    \centering
    \begin{tabular}{SSSSSS} \\ \toprule
    {$Experiment\_label$} & {$FW\_time$} & {$BW\_time$} & {$MP\_time$} & {$Time\_saving\_wrt\_FW$} & {$Time\_saving\_wrt\_BW$}\\ \midrule
    1  & 0.0012s & 0.0096s & 0.0003s & 76.19\% & 97.01\% \\
    2  & 0.0004s & 0.0070s & 0.0004s & -4.65\% & 94.29\% \\
    3  & 0.0026s & 0.0389s & 0.0023s & 13.79\% & 94.17\% \\
    4  & 0.0005s & 0.0127s & 0.0005s & 3.10\% & 96.11\% \\
    5  & 0.0002s & 0.0106s & 0.0001s & 14.94\% & 98.75\% \\
    6  & 0.0005s & 0.0020s & 0.0005s & 1.72\% & 74.65\% \\
    7  & 0.0004s & 0.0056s & 0.0003s & 8.29\% & 94.07\% \\
    8  & 0.0033s & 0.0429s & 0.0033s & 1.49\% & 92.32\% \\
    9  & 0.0001s & 0.0026s & 0.0001s & 11.52\% & 95.06\% \\
    10 & 0.0006s & 0.0011s & 0.0005s & 15.38\% & 50.55\% \\ \bottomrule \\
\end{tabular}
\caption{Third experiments battery. Number of nodes ranging between [100, 1000], target\_safe\_ratio ranging between [0.1,1.0].}
\label{tab:100_1000_nodes_nb}
\end{table*}

\begin{table*}[ht]
    \centering
    \begin{tabular}{SSSSSS} \toprule
    {$Experiment\_label$} & {$Total\_nodes$} & {$Total\_edges$} & {$Target\_nodes$} & {$Safety\_nodes$} & {$Reachability\_nodes$} \\ \midrule
    1  &5341 &9515 &1226 &4587 &754 \\
    2  &6368 &8771 &1037 &4883 &1485 \\
    3  &5887 &28147 &1887 &5452 &435 \\
    4  &5990 &8531 &148 &1111 &4879 \\
    5  &6275 &25183 &325 &1038 &5237 \\
    6  &5667 &20984 &2613 &2933 &2734 \\
    7  &5961 &23184 &1461 &5679 &282 \\
    8  &5543 &7214 &554 &3868 &1675 \\
    9  &5318 &16082 &2270 &1648 &3670 \\
    10  &5581 &20494 &1236 &61 &5520 \\ \bottomrule
\end{tabular}
    \centering
    \begin{tabular}{SSSSSS} \\ \toprule
    {$Experiment\_label$} & {$FW\_time$} & {$BW\_time$} & {$MP\_time$} & {$Time\_saving\_wrt\_FW$} & {$Time\_saving\_wrt\_BW$}\\ \midrule
    1  & 0.0494s & 2.8728s & 0.0409s & 17.06\% & 98.57\% \\
    2  & 0.0689s & 4.3557s & 0.0596s & 13.40\% & 98.63\% \\
    3  & 0.0479s & 3.7482s & 0.0438s & 8.68\% & 98.83\% \\
    4  & 0.0101s & 1.1958s & 0.0113s & -12.04\% & 99.06\% \\
    5  & 0.0196s & 1.6087s & 0.0177s & 9.72\% & 98.90\% \\
    6  & 0.1544s & 1.8811s & 0.1561s & -1.08\% & 91.70\% \\
    7  & 0.0275s & 3.9568s & 0.0231s & 16.22\% & 99.42\% \\
    8  & 0.0368s & 3.0642s & 0.0316s & 14.10\% & 98.97\% \\
    9  & 0.1412s & 1.3840s & 0.1292s & 8.48\% & 90.66\% \\
    10 & 12.5055s & 0.7117s & 0.0048s & 99.96\% & 99.33\% \\ \bottomrule \\
\end{tabular}
\caption{Fourth experiments battery. Number of nodes ranging between [5000, 6500], target\_safe\_ratio ranging between [0.1,1.0].}
\label{tab:5000_6000_nodes}
\end{table*}
\FloatBarrier
\end{samepage}

\bibliographystyle{unsrt}  
\bibliography{references}
\clearpage
\section*{Appendix: pseudocode}
\begin{OptAlgo}[H]
\caption{\label{alg:optAlgo}}
\begin{algorithmic}[1]
\Statex{\textbf{\newline Input:}}
    \Statex{$G:(V,\,E)$ s.t. V set of nodes, E set of edges bw nodes in V }
    \Statex{$V_S \subseteq V$: Set of nodes controlled by the safety player}
    \Statex{$V_R \subseteq V$: Set of nodes controlled by the reachablity player}
    \Statex{Target\_safe: Set of target nodes for the safety player}
    \Statex{Threshold: Threshold that regulates the strategy shift}
\Statex{\textbf{Output}:} 
    \Statex{Win: The winning set of nodes for the safety player} \newline
\Statex{\textbf{Begin: }}
    \State{Win $\gets$ Target\_safe}
    \State{Lose $\gets$ $V\;\backslash$ Target\_safe}
    \State{Last\_force\_reach $\gets$ Lose}
    \While{True}
        \If{$Card(Win)\footnotemark[6] \leq$  Threshold }
            \State{F $\gets$ STEP\_FORWARD($G=(V,\,E),\,Win$)}
            \State{Win\_new := Win $\cap$ F}
            \State{Win $\gets$ Win\_new}
            \State{Lose $\gets$ Lose $\cup$ [Win $\triangle$ F]}
        \EndIf
        \State{\textbf{else}} \hfill \text{//$Card(Win) >$ Threshold}
        \State{\qquad F$\gets$STEP\_BACKWARD(\newline $G,\,G^T,\,Lose,\, Last\_force\_reach$)}
        \State{\qquad Win $\gets$ Win $\setminus$ F}
        \State{\qquad Lose $\gets$ Lose $\cup$ F}
        \State{\qquad Last\_force\_reach $\gets$ F}
        \If{$Card(Win\_new) = Card(Win)$ \footnotemark[7]}
            \State{\textbf{return} Win}
        \EndIf
    \EndWhile
\newline
\Function{STEP\_FORWARD}{$G=(V,\,E),\,Win$}:
    \State{F\_s $\gets$ FORCE\_SAFE\_FORWARD($Win$)}
    \State{F\_r $\gets$ FORCE\_REACH\_FORWARD($Win \cap V_R$)}
    \State{F $\gets$ F\_s $\cup$ F\_r}
    \State{\textbf{return} F}
\EndFunction
\newline
\Function{STEP\_BACKWARD}
{\newline $G=(V,\,E),\,G^T,\,Lose,\,Last\_force\_reach$}:
    \State{F\_r $\gets$ FORCE\_REACH\_BACKWARD(\newline$G^T,\,Lose,\,Last\_force\_reach$)}
    \State{F\_s $\gets$ FORCE\_SAFE\_BACKWARD($G,\,G^T,\,Lose$)}
    \State{F $\gets$ F\_s $\cup$ F\_r}
    \State{\textbf{return} F}
\EndFunction
\end{algorithmic}
\end{OptAlgo}
\footnotetext[0]{\textsuperscript{6}The operator $Card(.)$ returns the cardinality of a given input set.}
\footnotetext[0]{\textsuperscript{7}For the sake of efficiency, we compare the sets' cardinalities instead of comparing their elements to verify if the fix-point has been reached.}
\footnotetext[0]{\textsuperscript{8}If a safe node is isolated, then it is clearly safe.}

\begin{Auxiliary_funs}[H]
\caption{\label{alg:optAlgoAux}}
\begin{algorithmic}[1]
\Function{FORCE\_SAFE\_FORWARD}{$\newline G=(V,\,E),\,Win$}:
    \State{F $\gets \emptyset$}
    \State{Win\_safety $\gets$ Win $\cap\;V_S:$}
    \For{u $\in$ Win\_safety}
        \If{$Card(Neighbors(u,G)) = 0$ \footnotemark[8]} 
        \State{F $\gets$ F $\cup\;\{u\}$}
        \EndIf
        \If{$(\exists\;u \in Neighbors(u,\,G)\,\vert\,v \in Win)$:}
            \State{F $\gets$ F $\cup\;\{u\}$}
        \EndIf
    \EndFor
    \State{\textbf{return} F}
\EndFunction
\newline
\Function{FORCE\_REACH\_FORWARD}{$\newline G=(V,\,E),\,Win$}:
    \State{F $\gets \emptyset$}
    \State{Win\_reach $\gets$ Win $\cap\;V_R:$}
    \For{u $\in$ Win\_reach:}
        \If{$Neighbors(u,\,G) \subseteq$ Win}
            \State{F $\gets$ F $\cup\;\{u\}$}
        \EndIf
    \EndFor
\State{\textbf{return} F}
\EndFunction
\newline

\Function{FORCE\_REACH\_BACKWARD}
{\newline$G^T=(V,\,E^T),\,V_R,\,Lose,\,Last\_force\_reach$}:
    \State{F $\gets \emptyset$}
    \For{u $\in$ Last\_force\_reach:}
        \State{F $\gets$ F $\cup\,$ ($Neighbors(u,\,G^T)\cap\,V_R$)}
    \EndFor
\State{\textbf{return} F}
\EndFunction
\newline

\Function{FORCE\_SAFE\_BACKWARD}{\newline $G^T=(V,\,E^T),\,Lose$}:
    \State{F $\gets \emptyset$}
    \State{Processed $\gets \emptyset$}
    \For{u $\in$ Lose:}
        \For{v $\in$ ($Neighbors(u,\,G^T)\cap\,V_S$)}
        \If{v $\notin$ (Processed $\cup$ Lose)}            
            \If{$Neighbors(v,\,G) \subseteq Lose$}
                \State{F $\gets$ F $\cup$ \{v\}}
            \EndIf
            \State{Processed $\gets$ Processed $\cup$ \{v\}}
        \EndIf
        \EndFor
    \EndFor
    \State{\textbf{return} F}
\EndFunction
\newline
\end{algorithmic}
\end{Auxiliary_funs}
\clearpage
\begin{NaiveForceReach}[H]
\begin{algorithmic}[1]
\caption{\newline Unoptimized version of the FORCE\_REACH function.}
\Function{NAIVE\_FORCE\_REACH}{$G^T, V_R, C$}
\Statex{\textbf{Input:}}
\Statex{$G^T:(V,\,E^T)$ s.t. V set of nodes, $E^T$ set of edges obtained by inverting the edges in G}
\Statex{$V_R$ $\subseteq$ V //Set of nodes controlled by the reachability player}
\Statex{Q $\subseteq$ V //Target set for the reachability player}
\Statex{\textbf{Output}:} 
\Statex{F $\subseteq$ V \hfill //F = \{v $\in$ $V_R\,:$ $\exists$ u $\in$ Q $\vert$ (u,v) $\in$ $E^T$\}}
\State{F $\gets$ $\emptyset$}
    \For{u $\in$ Q}
        \For{u $\in$ $V_R$}
            \If{u $\in$ \textit{Neighbors($u,\,G$)}}
                \State{F $\gets$ F $\cup$ \{v\}}
            \EndIf
        \EndFor
    \EndFor
\State{\textbf{return} F}
\EndFunction
\end{algorithmic}
\end{NaiveForceReach}
\begin{ReachGame}[H]
\caption{\label{alg:ReachAlgo}}
\begin{algorithmic}[1]
    \Statex{\textbf{\newline Input:}}
        \Statex{$G:(V,\,E)$ s.t. V set of nodes, E set of edges bw nodes in V }
        \Statex{T $\subseteq$ V: Target set for the reachability player}
        \Statex{$V_S,\,V_R$ := Sets of nodes $v \in V$ s.t. $V_S \cup V_R = V$} \footnotemark[4]
    \Statex{\textbf{Output}:} 
        \Statex{Win $\subseteq$ V: Winning set for the reachability player} \newline
    \Statex{\textbf{Begin: }}
        \State{$G^T=(V, E^T) \gets$ transpose(G)}
        \State{Q $\gets$ T}\hfill// Q:= Target set for the reachability player
        \State{C $\gets$ T}\hfill// C:= Set of nodes found reachable in the last iteration
        \State{F\textsubscript{reach} $\gets$ FORCE\_REACH($G^T,\,V_R,\,C$)}
        \State{F\textsubscript{safe} $\gets$ FORCE\_SAFE($G,\, G^T,\,V_S,\,Q$)}
        \State{F $\gets$ F\textsubscript{reach} $\cup$ F\textsubscript{safe}}
        \State{Q' $\gets$ Q $\cup$ F }
        \While{Q != Q':}
        \State{C $\gets$ F}
        \State{Q $\gets$ Q'}
        \State{F\textsubscript{reach} $\gets$ FORCE\_REACH($G^T,\,V_R,\,C$)}
        \State{F\textsubscript{safe} $\gets$ FORCE\_SAFE($G,\, G^T,\,V_S,\,Q$)}
        \State{F $\gets$ F\textsubscript{reach} $\cup$ F\textsubscript{safe}}
        \EndWhile
        \State{\textbf{return Q}}
    \State{\textbf{End}}
\end{algorithmic}
\end{ReachGame}
\begin{Improved_algo_auxiliary_funs}[H]
\begin{algorithmic}[1]
\caption{}
\Function{FORCE\_REACH}{$G^T,\,V_R,\,C$}
\Statex{\textbf{Input:}}
\Statex{$G^T:(V,E^T)$ s.t. V set of nodes, $E^T$ set of edges obtained by inverting the edges in G}
\Statex{$V_R \subseteq$ V //Set of nodes controlled by the reachability player}
\Statex{C $\subseteq$ V //Set of nodes found reachable in the last iteration}
\Statex{\textbf{Output}:} 

\Statex{F $\subseteq$ V \hfill //F = \{v $\in$ $V_R$: $\exists$ u $\in$ C $\vert$ (u,v) $\in$ $E^T$\}} \footnotemark[5]
\State{F $\gets$ $\emptyset$}
    \For{u $\in$ C}
    \State{F $\gets$ F $\cup$ (\textit{Neighbors(u, $G^T \cap V_R$))}}  
    \EndFor
\State{\textbf{return} F}
\EndFunction
\newline
\Function{FORCE\_SAFE}{$G,\,G^T,\,V_S,\,Q$}
\Statex{\textbf{Input:}}
\Statex{$G:(V,\,E)$ s.t. V set of nodes, E set of edges bw nodes in V }
\Statex{$G^T:(V,E^T)$ s.t. V set of nodes, $E^T$ set of edges obtained by inverting the edges in G}
\Statex{$V_S$ $\subseteq$ V //Set of nodes controlled by the safety player}
\Statex{Q $\subseteq$ V //Target set for the reachability player}
\Statex{\textbf{Output}:}
\Statex{F $\subseteq$ V \hfill //F = \{v $\in$ $V_S$ $\vert$ \{ $u\,\vert \,(v,\,u)$ $\in$ E \} $\subseteq$ Q \}} \footnotemark[6]
    \For{u $\in$ Q}
        \State{N $\gets$ \textit{Neighbors($u,\,G^T$)} $\cap$ $V_S$}
        \For{v $\in$ N:}
            \If{\textit{Neighbors($v,\,G$)} $\subseteq$ Q}
            \State{F $\gets$ F $\cup$ \{v\}}
            \EndIf
        \EndFor
    \EndFor
    \State{\textbf{return} F}
\EndFunction
\end{algorithmic}
\end{Improved_algo_auxiliary_funs}
\footnotetext[0]{\textsuperscript{4}We define $V_S$ as the set of nodes controlled by the safety player, while we indicate with $V_R$ the set of nodes controlled by the reachability player.}
\footnotetext[0]{\textsuperscript{5}We indicate with F all the nodes controlled by the reachability player that have an edge to a node in C. Please notice we employ the transpose graph as an optimization.}
\footnotetext[0]{\textsuperscript{6}We indicate with F all the nodes controlled by the safety player that have an edge to a node in Q. Please notice we employ the direct graph in this case.}


\end{document}